\documentclass[aps,showpacs,pra,floatfix,reprint,superscriptaddress]{revtex4-1}
\usepackage{graphicx}
\usepackage{bm,epsfig}
\usepackage{amsmath}
\usepackage{amssymb}
\usepackage{natbib}
\usepackage{color,enumerate}

\begin{document}

\title{Numerical Optical Centroid Measurements}

\author{Qurrat-ul-Ain \surname{Gulfam}}
\email{qurrat-ul-ain@mpi-hd.mpg.de}
\affiliation{Max-Planck-Institut f\"ur Kernphysik, Saupfercheckweg 1, 
D-69117 Heidelberg, Germany}

\author{J\"org \surname{Evers}}
\email{joerg.evers@mpi-hd.mpg.de}
\affiliation{Max-Planck-Institut f\"ur Kernphysik, Saupfercheckweg 1, 
D-69117 Heidelberg, Germany}


\date{\today}

\begin{abstract}
Optical imaging methods are typically restricted to a resolution of order of the probing light wavelength $\lambda_p$ by the Rayleigh diffraction limit. This limit can be circumvented by making use of multiphoton detection of correlated $N$-photon states, having an effective wavelength $\lambda_p/N$. But the required $N$-photon detection usually renders these schemes impractical. To overcome this limitation, recently, so-called optical centroid measurements (OCM) have been proposed which replace the multi-photon detectors by an array of single-photon detectors. Complementary to the existing approximate analytical results, we explore the approach using numerical experiments by sampling and analyzing detection events from the initial state wave function. This allows us to quantitatively study the approach also beyond the constraints set by the approximate analytical treatment, to compare different detection strategies, and to analyze other classes of input states.
\end{abstract}


\maketitle

\section{Introduction}

Conventional optical lithography schemes are impeded by the Rayleigh diffraction limit~\cite{Rayleigh,abbe} owing to the wave nature of light. Not least due to the continuous and increasing demand of faster devices in microelectronics, smaller structures in semiconductors need to be written. But highly energetic photons can lead to enhanced material damage and demand the development of new optical imaging systems. Motivated by this, several alternative routes have been explored.
A prominent example is quantum lithography~\cite{boto,bjoerk,angelo,ql-review}  based on number-path entangled states of light. But generation of the required entangled states, decoherence due to photon loss and detection remain a challenge. 
Alternatively, classical light fields can be used. One tool is post-selection of certain detection events which contain image data with high resolution~\cite{retrodictive,PhysRevLett.105.163602}. 
Other methods aim at facilitating a $N$-photon absorption process leading to a reduced  effective wave length~\cite{vrijin,classical1,Friesem,nphot,classical2}. These, however,  require multiphoton transitions, which complicate a practical implementation. This requirement can be relaxed by means of dark states in multi-level systems~\cite{kiffner} or of Rabi oscillations in two-level systems~\cite{PhysRevLett.105.183601}.

Recently, a different method has been proposed to remove the requirement of multi-photon detection~\cite{centroid}. In the so-called optical centroid method (OCM), the $N$-photon detector is replaced by an array of single-photon detectors. For each input pulse, the results of the individual single-photon detectors are then combined into a single variable, the centroid. The probability distribution of the centroid obtained from many repetitions of the measurement then contains the desired sub-wavelength resolution. Thus, no $N$-photon absorber is required, and at the same time, the detection becomes more efficient as all events rather than only $N$-photon events contribute to the signal. Recently, a proof-of-principle experiment of OCM for two photon entangled state has been carried out by~\cite{PhysRevLett.107.083603}. The experiment used two kinds of detectors: One is a pair of single photon detectors for OCM while the other is two-photon absorption detector for QL. OCM achieved the same quantum 
superresolution as QL but much more efficiently.

Motivated by this, here, we study the OCM using numerical experiments. In contrast to existing approximate analytical approaches, the numerical study allows us to quantitatively study OCM over a broader parameter range, to compare different detection strategies, and to analyze other classes of input states. In particular, in addition to the NOON and jointly Gaussian input states proposed in~\cite{centroid}, we analyze cat states made out of superpositions of coherent states. 
For our analysis, we first generate random detection events from position space  wave function. Afterwards, we model the detection by  a suitable discretization using different methods, and apply the centroid method. We characterize the error in the recovery of the theoretical centroid distribution by root mean square (rms) deviation from the original wave function. 
We confirm our method by recovering key predictions from the analytical calculations~\cite{centroid} quantitatively, such as the dependence of multiphoton count rates for jointly Gaussian states on the properties of the input state.

We then study the dependence of the rms deviation on the size of the individual detectors in the array. We find several features which may appear counter-intuitive at first sight, but can be explained in terms of the measurement procedure chosen and of statistical effects most pronounced in the small detector limit. For intermediate detector sizes, the rms deviation is nearly a linear function of the detector size. We analyze this dependence for NOON-states with different $N$ and show how the slope of the linear dependence can be traced back to the structure of the underlying wave function. We then extract single-and multi-photon detection events, and verify the improved efficiency expected for the OCM over $N$-photon absorption schemes. For the jointly Gaussian state, we augment this analysis by studying the rms deviation for different $N$ as a function of the detector size while keeping the feature size constant. We finally turn to coherent cat states, and show that the OCM can also be applied to them. 
These states are of interest, as the feature size of the cat states can continuously be tuned via the coherent state parameter $\alpha$. Furthermore, the structure of the cat state wave function along the centroid axis can be modified via the phase of $\alpha$ such that it differs from the regular patterns observed for NOON or Gaussian states.

The paper is organized as follows. In Sec.~\ref{theory}, we provide the theoretical background. In particular, we review the OCM method in~\ref{ocm-theory}, introduce the various non-classical input states in Sec.~\ref{trial-states}, discuss the implementation of the numerical experiments (Sec.~\ref{numerical-exp}) and the detection system (Sec.~\ref{discretization}), and describe the procedure to estimate the error in the OCM recovery in Sec.~\ref{error-rms}. In Sec.~\ref{results}, our results for different types of non-classical input states are discussed. The final Sec.~\ref{summary} summarizes the results.

%

\section{Theoretical considerations}
\label{theory}

\subsection{Centroid method}\label{ocm-theory}

We briefly review the formalism of resolution enhancement developed in~\cite{multiphoton,centroid} using quantum lithography and optical centroid measurements. We consider photons of wavelength $\lambda_p$ incident on a detection plane under an angle such that the wavelength associated to the wave vector component in the detection plane is given by $\lambda = \lambda_p / \sin \theta$. In the following, we restrict the analysis to one spatial dimension in the detection plane. Let $\hat{a}(k)(\hat{a}^\dag(k))$ denote the photon annihilation (creation) operator in this transverse momentum space. 

These operators follow the commutation relation
\begin{equation}
 [\hat{a}(k),\hat{a}^{\dag}(k')]=\delta(k-k')\,.
\end{equation}
An $N$-photon momentum eigenvector can be defined using $\hat{a}^\dag(k_i)$ as
\begin{equation}
 |k_1,k_2,\dots,k_N\rangle = \frac{1}{\sqrt{N !}}\hat{a}^\dag(k_1)\hat{a}^\dag(k_2)\dots\hat{a}^\dag(k_N)|0\rangle\,,
\end{equation}
where $|0\rangle$ indicates the state without photons.
The momentum space wave function representation of a pure $N$-photon Fock state $|N\rangle$ is given by
\begin{equation}
 \phi(k_1,k_2,\dots,k_N)=\langle k_1,k_2,\dots,k_N|N\rangle. 
\end{equation}
The transverse momenta of the photons are restricted by the Rayleigh diffraction limit~\cite{Rayleigh} such that
\begin{equation}
\phi(k_1,k_2,\dots,k_N)=0 
\end{equation}
for any $|k_n|>2\pi/\lambda$. 
The spatial annihilation operator is defined as
\begin{equation}
 \hat{A}(x)\equiv \frac{1}{\sqrt{2 \pi }}\int dk \hat{a}(k)\,e^{i k x}\,.
\end{equation}
A corresponding $N$-photon state can be constructed as
\begin{equation}
 |x_1,x_2,\dots,x_N\rangle \equiv \frac{1}{\sqrt{N !}}\hat{A}^\dag(x_1)\dots\hat{A}^\dag(x_N)|0\rangle\,.
\end{equation}
The position space wave function is obtained by taking the $N$-dimensional Fourier transform of $\phi(k_1,k_2,\dots,k_N)$ and is given by
\begin{equation}
\psi(x_1,x_2,\dots,x_N)=\langle x_1,x_2,\dots,x_N|N\rangle.
\end{equation}
Of fundamental importance in the centroid scheme is the transformation of the photon position coordinates $x_i$ to the centroid coordinate $X$ and the relative position coordinates $\xi_n$ which are defined as
\begin{equation}
\label{centroid-def}
 X\equiv \frac{1}{N}\sum_{n=1}^N x_n, \qquad \xi_n\equiv x_n-X.
\end{equation}
Note that when $x_1=x_2=\dots=x_N=x$, $X=x$, that is, if all photons are constrained to arrive at the same point $x$, the relative positions $\xi_n$ become zero and the centroid variable also lies at the same point $x$ according to Eq.~(\ref{centroid-def}).

Conventional quantum lithography proposals rely on $N$-photon detection at a single detector at position $x$. The input state is  a $N$-photon state $\hat{\rho}_N=|N\rangle\langle N|$. The probability density for multi photon absorption is therefore given by
\begin{align}\label{dm}
p_{MP}(x)=\langle:\hat{I}^N(x):\rangle&=\langle [\hat{A}^{\dag}(x)]^N[\hat{A}(x)]^N\rangle\nonumber\\&=N|\psi(x,x,\dots,x)|^2.
\end{align}

The centroid method, in contrast, relies on an array of detectors, such that each of the incident $N$ photons is detected by an individual detector. Out of the individual photon detection positions, the centroid position $X$ is calculated. The probability density of the centroid distribution is obtained from 
\begin{align}
\langle:\Pi_{n=1}^N\hat{I}(x_n):\rangle
\end{align}
as a marginal probability density by tracing out all relative positions $\xi_n$ using $x_n = X + \xi_n$,
\begin{align}\label{dc}
&p_{OCM}(X)=\int d\xi_1,\dots,d\xi_{N-1}\langle:\Pi_{n=1}^N\hat{I}(X+\xi_n):\rangle\nonumber\\
&=\int d\xi_1,\dots,d\xi_{N-1}N|\psi(X+\xi_1,\dots,X+\xi_N)|^2.
\end{align}
The key result of Tsang's analysis is that the two distributions Eq.~(\ref{dc}) and Eq.~(\ref{dm}) have the same spatial dependence and thus resolution.
Therefore, the technically challenging $N$-photon detection in a single detector can be replaced by $N$ single-photon detections. A further advantage of the centroid method is that it is potentially much more efficient than multi-photon detection, i.e., for experimentally relevant parameters, $p_{OCM}(X)$ can exceed $p_{MP}(X)$ significantly~\cite{multiphoton,centroid,PhysRevLett.107.083603}.

\subsection{Trial states}\label{trial-states}

Throughout our analysis, we apply the optical centroid method using numerical experiments to various non-classical photon states, which we introduce in the following.

\subsubsection{NOON states}
\label{noon-sec}

Following~\cite{multiphoton}, we consider momentum-correlated $N$-photon NOON states 
\begin{align}
|NOON\rangle = \frac{1}{\sqrt{2}}\left( |N\rangle_A |0\rangle_B +  |0\rangle_A |N\rangle_B \right )
\end{align}
in which the two modes $A$ and $B$ correspond to two wave vector directions for the photons incident on the detection plane, with mean transverse component of the  wave vector $k_0$ and $-k_0$, respectively. The momentum spread of the transverse wave vector is denoted by $\Delta k$ with $k_0 \gg \Delta k$. Denoting the (normalized and even) transverse wave vector profile as $f(q)$, the momentum space wave function is
\begin{align}
\label{phinoon}
 \phi_{NOON}(k_1, k_2, \dots, k_N)&= \frac{1}{\sqrt{2\Delta k^N}}\left [ \prod_{n=1}^N f\left (\frac{k_n-k_0}{\Delta k}\right ) \right.\nonumber \\
&
\left. +\prod_{n=1}^N f\left (-\frac{k_n-k_0}{\Delta k} \right)\right ]\,.
\end{align}
By Fourier transform, assuming $f(q)=\pi^{-1/4} e^{-q^2/(2\Delta k^2)}$, we then obtain the position space wave function
\begin{align}
\psi_{NOON}(x_1,x_2,\dots,x_N) = &\frac{ \sqrt{2 \, \Delta k^N}}{\pi^{N/4}} \: e^{-\frac{1}{2}\Delta k^2\: \sum_{i=1}^N x_i^2}  \nonumber \\
&  \times \cos\left ( k_0 \sum_{i=1}^{N} x_i \right )\,.
\end{align}

The probability distribution thus follows as
\begin{align}
\label{prob-noon}
 &|\psi_{NOON}(x_1,x_2,\dots,x_N)|^2
\nonumber \\
& =2\, \left( \frac{ \Delta k}{\sqrt{\pi}}\right)^N \: e^{-\Delta  k^2\: \sum_{i=1}^N x_i^2}  
\, \cos^2\left ( k_0 \sum_{i=1}^{N} x_i \right )\,.
\end{align}
Note that this expression is only normalized in the limit $\Delta k \ll k_0$, in which the field operators for the two modes $A$ and $B$ satisfy the usual bosonic commutation rules.

For numerical simulations we employ dimensionless position variables $\bar{x}=x/ \lambda$ such that the position coordinates are measured in the units of $\lambda$. In terms of the dimensionless variables, Eq.~(\ref{prob-noon}) becomes
\begin{align}
2\, \left( \frac{ \Delta k}{\sqrt{\pi}}\right)^N \: e^{-\frac{4\pi^2}{\sigma^2}\: \sum_{i=1}^N {\bar x}_i^2}  
\, \cos^2\left ( 2\pi\: \sum_{i=1}^{N} \bar{x}_i \right )\,.
\end{align}
Using $\lambda = 2\pi/k_0$, we have defined the dimensionless width $\sigma=k_0/\Delta k$. Throughout the numerical analysis, we have chosen $\sigma=4\sqrt{2}\pi$ such that $\Delta k \ll k_0$ as required.

\begin{figure}[t!]
\includegraphics[width=0.48\linewidth ]{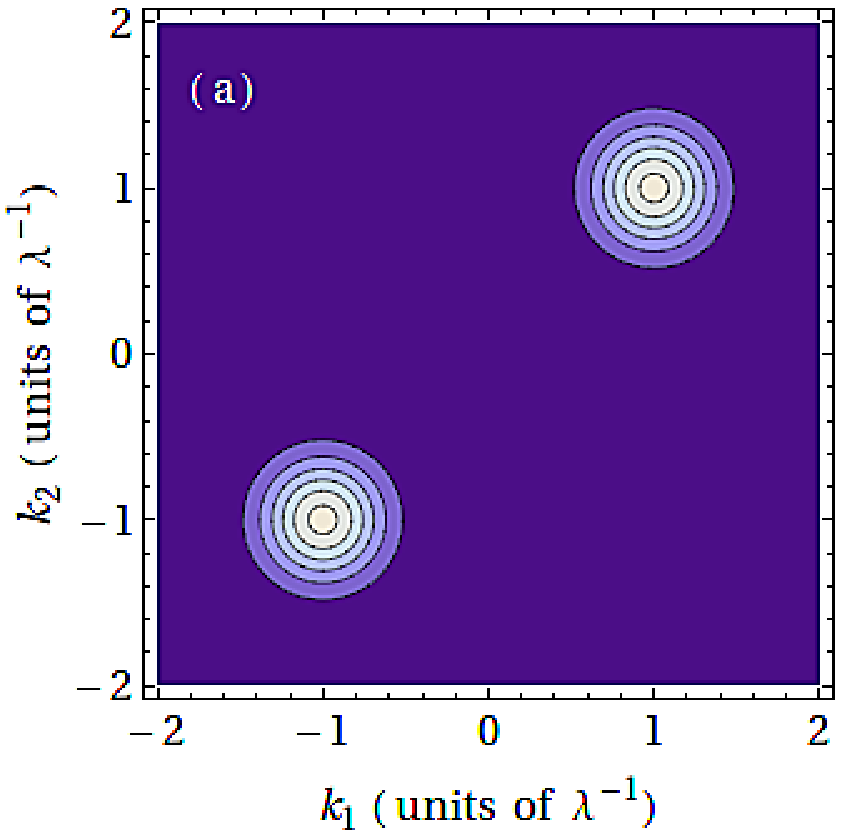}
\includegraphics[width=0.48\linewidth ]{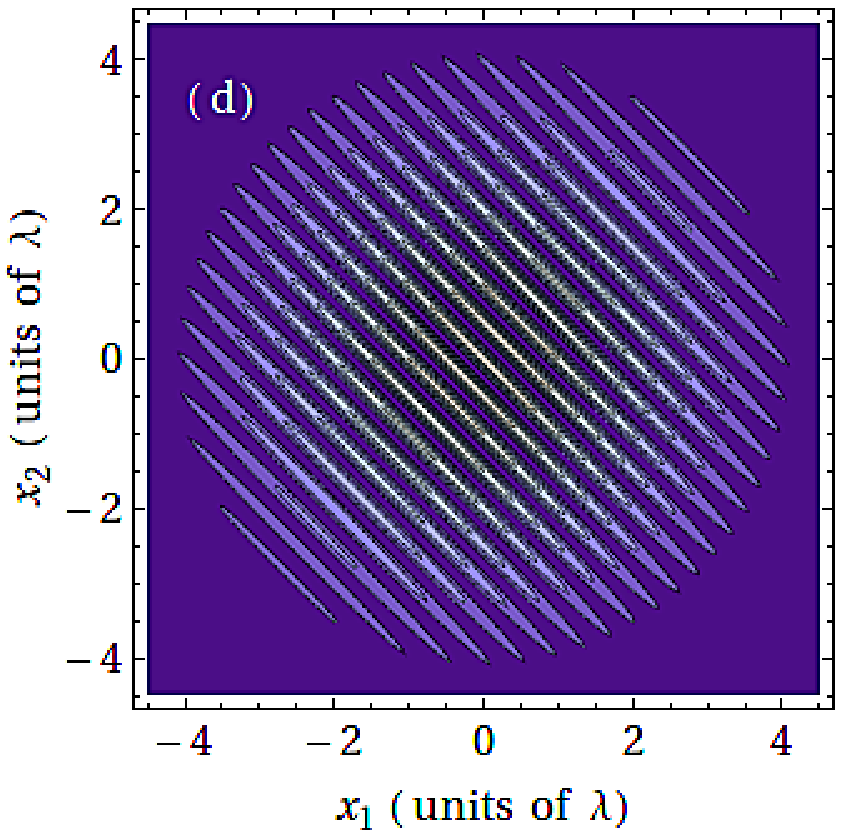}\\
\includegraphics[width=0.48\linewidth ]{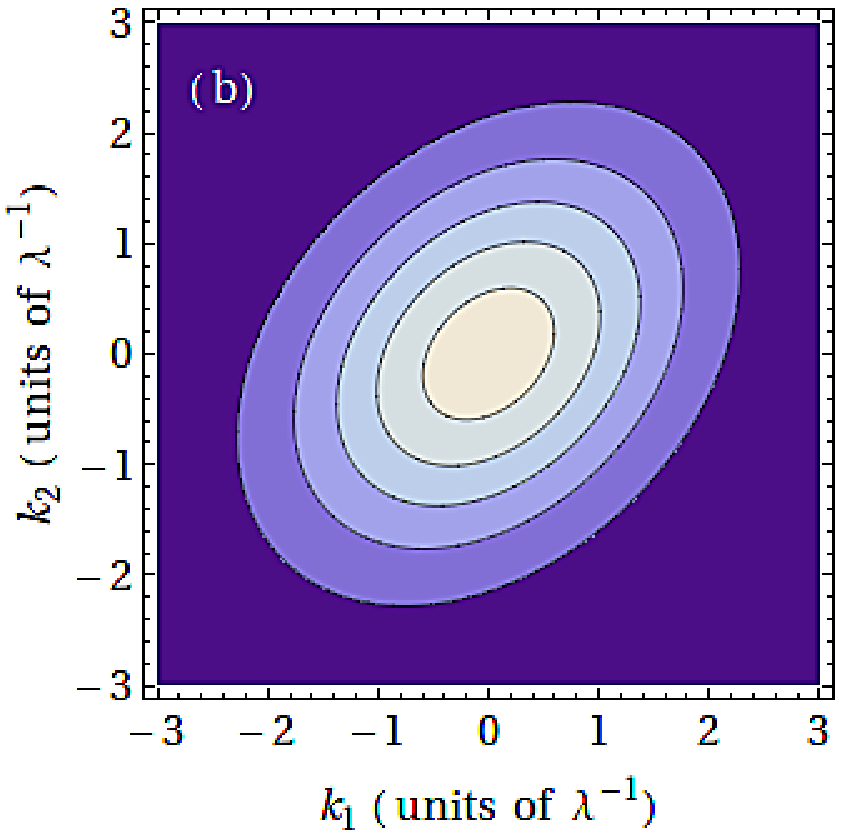}
\includegraphics[width=0.48\linewidth ]{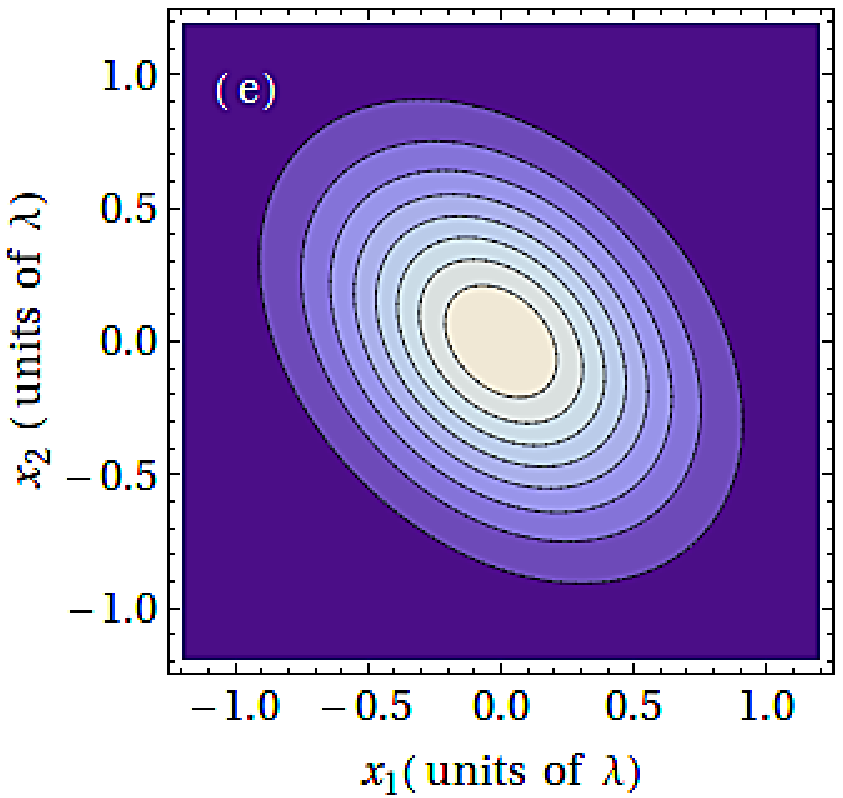}\\
\includegraphics[width=0.48\linewidth ]{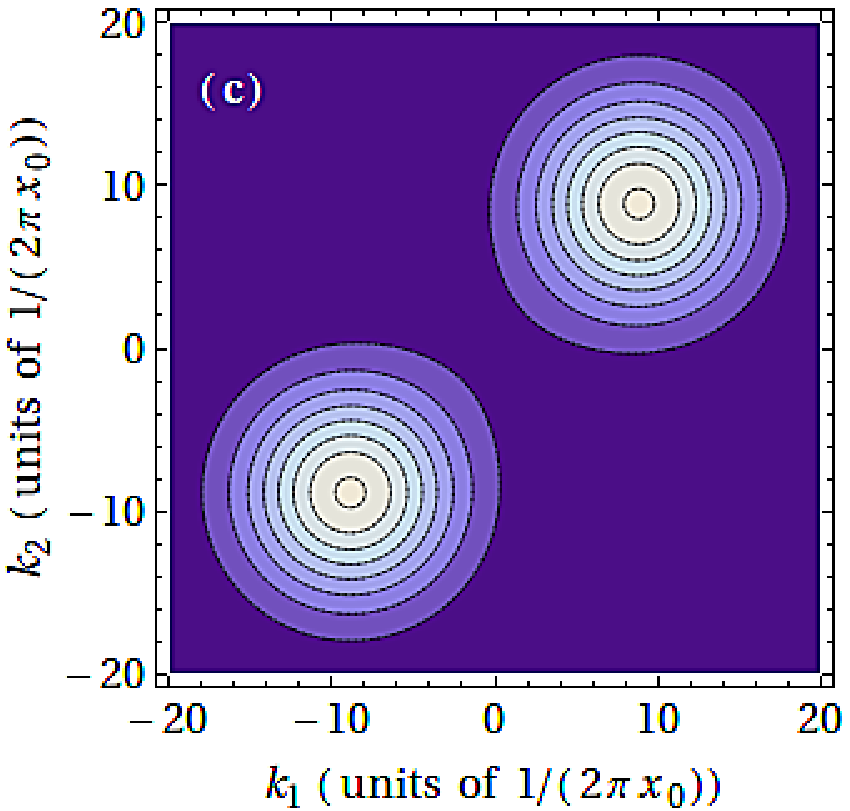}
\includegraphics[width=0.48\linewidth ]{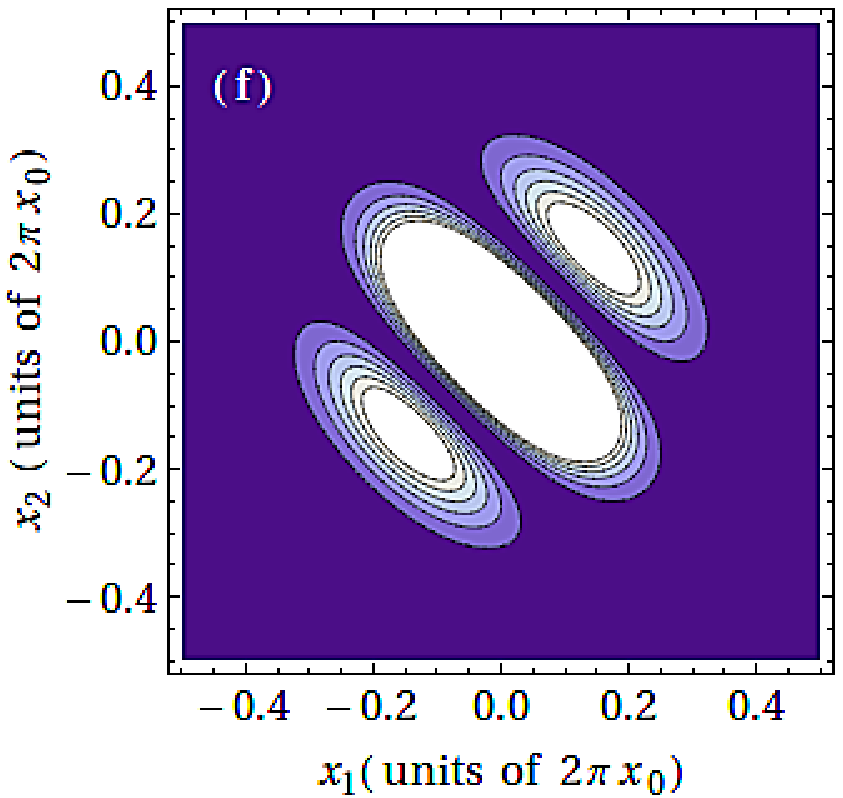}\\

\caption{\label{fig-states}(Color online) Momentum (a-c) and the corresponding position (d-f) space probability distributions of a two-photon NOON state (top row), joint Gaussian state with $N=2$ and $B=\beta=\lambda^{-1}$ (middle row), and correlated cat state with $\alpha=i$ (bottom row). }
\end{figure}

An example for the position space probability distribution of a two-photon NOON state using dimensionless variables is shown in Fig.~(\ref{fig-states}).

2-photon NOON states can be created using spontaneous parametric down conversion combined with the Hong-Ou-Mandel effect~\cite{boto,HOM}. Such NOON states have been employed in proof-of-principle experiments on resolution enhancement~\cite{angelo,Kawabe:07,PhysRevLett.107.083603}. Higher order NOON states with larger $N$ have also been created experimentally~\cite{Nagata04052007,mitchell,
walther,afek,*afek2,six}.

\subsubsection{Jointly Gaussian states}
The momentum space representation of a jointly Gaussian state can be written as~\cite{multiphoton}
\begin{align}
\phi_{JG}(k_1, k_2, \dots, k_N) = \sqrt{\frac{C}{N}}\,e^{-\frac{1}{4}\sum_{n,m} k_n B_{nm}k_m}\,,
\end{align}
with components of the matrix $B$ given by
\begin{align}
 B_{nn}&=\frac{1}{N^2B^2}+\bigg(1-\frac{1}{N}\bigg)\frac{1}{\beta^2}\nonumber\,,\\
 B_{nm}&=\frac{1}{N^2B^2}-\frac{1}{N\beta^2}\quad (n\neq m)\,.
\end{align}
$C$ is a  normalization constant. $B$ can be interpreted as the width of the average momentum of the photons, whereas $\beta$ characterizes the width in momenta relative to the average~\cite{multiphoton}. 
These two parameters determine the variance of $k_n$ given by
\begin{equation}
\label{knsq}
 \langle k_n^2\rangle=B^2+(1-\frac{1}{N})\beta^2.
\end{equation}
The position wave function for the jointly Gaussian state then follows as
\begin{equation}
\label{jtg}
 |\psi_{JG}(x_1,x_2,\dots,x_N)\rangle\propto \exp(-\sum_{n,m}x_nB_{nm}^{-1}x_m)\,.
\end{equation}

The multi photon absorption pattern $\langle:\hat{I}^N(x):\rangle$ is a Gaussian with a root mean square width $1/(2NB)$.
If $B=\beta/\sqrt{N}$, the distribution becomes classical and the photons are independent. This distribution is a symmetric Gaussian with equal widths along the diagonal axes in the configuration space. The classical multi photon absorption width is given by the standard quantum limit $W_C=1/(2\sqrt{N\langle k_n^2\rangle})$. Increasing $B$ from the classical value makes the distribution quantum mechanical. The maximum value of $B$, $B=\sqrt{\langle k_n^2\rangle}$ renders the ultimate quantum limit for the width of the multi photon absorption pattern $W_{\rm{min}}=1/(2N\sqrt{\langle k_n^2\rangle})$. 

The quantum mechanical elongated jointly Gaussian has a shorter width along the centroid direction and is stretched along all the other diagonals in the position space. 

The position space probability distribution is
\begin{equation}
\label{prob-jg}
 |\psi_{JG}(x_1,x_2,\dots,x_N)\rangle|^2\propto \exp(-2\sum_{n,m}x_nB_{nm}^{-1}x_m)\,.
\end{equation}

Fig.~(\ref{fig-states}) shows the position space probability distribution for a quantum mechanical jointly Gaussian state in terms of dimensionless position variables defined in Sec.~(\ref{noon-sec}).

\subsubsection{Cat states}

The third class of states we investigate are correlated coherent cat states. We in particular analyze two mode states of the form~\cite{haroche}
\begin{equation}
\label{psi-ccc}
 |\psi_{ccc}\rangle=\mathcal{N}(|\alpha\rangle_1 |\alpha\rangle_2+|-\alpha\rangle_1 |-\alpha\rangle_2)\,,
\end{equation}
in which $|\alpha\rangle_i$ is a coherent state in mode $i$ with complex parameter $\alpha$, defined by $a_i |\alpha\rangle_i = \alpha |\alpha\rangle_i$ where $a_i$ is the photon annihilation operator of mode $i$. Note that other two mode cat states have also been considered in the literature in different contexts~\cite{othercat,PhysRevLett.91.230405,0953-4075-40-14-013}. These states can be seen as an extension of the famous single particle Schr\"odinger coherent cat states~\cite{cat+} given by $|\alpha_+\rangle\equiv|\alpha\rangle+|-\alpha\rangle$~\cite{coherent,*coherent2,*coherent3}. A proposal for the generation of Schr\"odinger's cat state has been discussed in~\cite{PhysRevA.59.4095}. Experiments performed on the optical cat states include~\cite{Ourjoumtsev07042006,PhysRevLett.97.083604,Wakui:07}. $\mathcal{N}$ is a normalization factor which is given by $1/\sqrt{2(1+\exp(-4|\alpha|^2)}$.

While Eq.~(\ref{psi-ccc}) resembles a two-photon NOON state, there is an important difference between the states. The bi photon NOON state consists of exactly two photons. Individual realizations of the same cat state, however, can contain different numbers of photons. The probability $p(n_1,n_2)$ to detect $n_i$ photons in mode $i$ evaluates to
\begin{align}
\label{pn1n2}
 &p(n_1,n_2)= \langle n_1,n_2|\psi_{ccc}\rangle\langle \psi_{ccc}|n_1,n_2\rangle\nonumber\\
=&2\mathcal{N}^2\frac{e^{-2|\alpha|^2}}{n_1!\: n_2!}|\alpha|^{2(n_1+n_2)}\:[1+(-1)^{n_1+n_2}].
\end{align}
It follows that the state can only contain even numbers of photons. For a certain range of $|\alpha|$, states with less or equal to two photons are populated with high probability. This motivates the analysis of detection events, in which one photon is detected in each of the two modes. In the following, we will focus on this case.

The position representation of a single mode coherent state is~\cite{schleich}
\begin{align}
\langle x|\alpha\rangle = \frac{1}{\sqrt{x_0\sqrt{\pi}}} e^{-\frac{1}{2}(\alpha^2-|\alpha|^2)}
e^{-\frac{1}{2}(x/x_0-\sqrt{2}\alpha)^2}
\end{align}
with $x_0 =\sqrt{\hbar/(m\omega_0)}$.
The position space representation of Eq.~(\ref{psi-ccc}) then follows as
\begin{align}
\label{xpsi-ccc}
 \psi_{ccc}(x_1,x_2)=& \langle x_1|\alpha\rangle_1 \langle x_2|\alpha\rangle_2+\langle x_1|-\alpha\rangle_1 \langle x_2|-\alpha\rangle_2\nonumber\\
=&\frac{2\mathcal{N}}{\sqrt{\pi}x_0} e^{-\frac{1}{2x_0^2}(x_1^2+x_2^2)}e^{-2|\alpha|\alpha \cos(\phi)}\nonumber \\
&\times \cosh\left(\frac{\sqrt{2}\alpha}{x_0}(x_1+x_2)\right)\,,
\end{align}
where $\alpha = |\alpha|\exp(i\phi)$.

For the further discussion, we now specialize to the case of $\alpha = i|\alpha|$ (i.e., $\phi=\pi/2$) to obtain a probability density
\begin{align}\label{psisq-cat}
 &|\psi_{ccc}(x_1,x_2,\alpha=i|\alpha|)|^2 \nonumber\\
&\quad =\frac{4\mathcal{N}^2}{\pi x_0^2} e^{-\frac{1}{x_0^2}(x_1^2+x_2^2)}
  \cos^2\left(\frac{\sqrt{2}|\alpha|}{x_0}(x_1+x_2)\right)\,.
\end{align}
The structure of this expression resembles the probability density for the two photon NOON states. Other choices for $\phi$ allow to rotate the fringe pattern in the position space with respect to the centroid axis. This way, the condition for the centroid method that the probability density should separate into a function of the centroid coordinate times a function of all other coordinates can be continuously violated by modifying $\phi$ from $\pi/2$.

As before, for the numerical calculations we define dimensionless position variables $\bar{x}$. For $x = 2\pi\,x_0\,\bar{x}$, Eq.~(\ref{psisq-cat}) becomes
\begin{align}\label{psisq-cat-scale}
 &|\psi_{ccc}(x_1,x_2,\alpha=i|\alpha|)|^2 \nonumber\\
&\quad =\frac{4\mathcal{N}^2}{\pi x_0^2} e^{-4\pi^2(\bar{x}_1^2+\bar{x}_2^2)}
  \cos^2\left(2\pi\,\sqrt{2}|\alpha|(\bar{x}_1+\bar{x}_2)\right)\,.
\end{align}
Comparing Eq.~(\ref{psisq-cat-scale}) with Eq.~(\ref{prob-noon}), we find that the exponential parts of the two expressions agree for $\sigma = 1$, whereas the cosine-parts differ in the sense that the cat state allows to continuously tune the structure size of the fringe pattern by varying $|\alpha|$. This way, the sub-wavelength resolution capabilities of the centroid method can be probed independent of the number of detected photons.

Note that alternatively, one could choose dimensionless variables $x = \sqrt{2}\pi\,x_0 \bar{x}/|\alpha|$, such that the cosine-parts of the NOON-state and the cat-state probability densities become the same. However, in this scaling, neither the overall scaling nor the width $\sigma$ of the cat state are independent of the choice of $|\alpha|$. 

The probabilty distribution for $\phi=\pi/2$ in configuration space is shown in Fig.~(\ref{fig-states}). Increasing $\phi$ from $\pi/2$ (Fig.~(\ref{fig-states}f) results in an increase of the width of each fringe along the centroid direction such that the fringes merge into each other. Further increase causes the middle fringe to become smaller and smaller until eventually at $\phi\approx \pi$, the two modes of the coherent states no more interfere and can be seen as separate circles.

\subsection{Numerical experiments}\label{numerical-exp}
In order to simulate the centroid method numerically, we first generate a large number of random photon events distributed and correlated according to the respective position space wave function. Next, we apply a discretization scheme to model the experimentally accessible signal for different detector characteristics. Finally, we perform the centroid analysis on this data to recover a wave function. We then compare this wave function to the original position space wave function and calculate the root-mean-square deviation to access the predictive power of the centroid method for the given conditions. In the following, we describe all steps in more detail.

\subsubsection{Random number generation}
\label{multivariates}
The first step is to generate random photon events according to the respective position space wave function. As can be seen from Eqs.~(\ref{prob-noon}), (\ref{prob-jg}) and (\ref{psisq-cat-scale}), the position variables are correlated. To facilitate the random event generation, we apply unitary transformations to a set of uncorrelated variables. For example, the probability distribution
\begin{align}
\label{p2}
2\, \left( \frac{ \Delta k}{\sqrt{\pi}}\right)^2 \: e^{-\Delta k^2\: (x_1^2 + x_2^2)}  
\, \cos^2\left [ k_0 (x_1 + x_2)  \right ]\,.
\end{align}
of Eq.~(\ref{prob-noon}) specialized to the case of two photons ($N=2$) does not allow to draw random positions $x_1, x_2$ independently. Applying the transformation 
\begin{equation}
 \vec{Y}_2=M_2 \vec{X}_2,
\end{equation}
with $\vec{X}_2 = (x_1, x_2)^T$,  $\vec{Y}_2 = (y_1, y_2)^T$, and
\[
M_2 =
\frac{1}{\sqrt{2}}\left[  {\begin{array}{cc}
 1 & -1  \\
 1 &  1  \\
 \end{array} } \right]\,,
\]
the probability distribution Eq.~(\ref{p2}) becomes
\begin{align}
2\, \left( \frac{ \Delta k}{\sqrt{\pi}}\right)^2 \: 
\left( e^{-\Delta  k^2\: y_1^2} \right)
\left( e^{-\Delta k^2\: y_2^2} \: \cos^2\left [\sqrt{2} k_0 y_2  \right ] \right)\,.
\end{align}
In this form, random values can be drawn for $y_1$ and $y_2$ independently. In a similar way, also 3- and 4-photon events can be handled, with coordinate transforms 
\[
M_3 =
\frac{1}{\sqrt{3}}\left[ {\begin{array}{ccc}
 1 & 1 & 1  \\
 0          & \sqrt{3/2} & -\sqrt{3/2} \\
 \sqrt{2} & -1/\sqrt{2} & -1/\sqrt{2} \\
 \end{array} } \right]
\]
and
\[
M_4 =
\frac{1}{\sqrt{4}}\left[ {\begin{array}{cccc}
 1 & 1 & 1 & 1  \\
 0   &  0  & \sqrt{2} & -\sqrt{2} \\
 0   &  2\sqrt{2/3} & \sqrt{2/3} & -\sqrt{2/3} \\
 \sqrt{6} & -1/\sqrt{3} & -1/\sqrt{3} & -1/\sqrt{3} \\
 \end{array} } \right]\,,
\]
respectively.

After application of the variable transform, for the NOON and correlated coherent cat states, $N-1$ of the new variables follow univariate normal distributions. The only exception is one variable which coincides with the centroid coordinate except for an overall scaling factor of $\sqrt{N}$. For the random number generation in this variable, we apply the cumulative distribution function. This function for a continuous random variable $V$ is obtained from the probability density function $\mathcal{I}$ via
\begin{equation}
 \mathcal{F}_V(v)=\int_{-\infty}^v\mathcal{I}(t) dt.
\end{equation}
If $x_i$ are random numbers drawn from the cumulative distribution, $\mathcal{F}^{-1}(x_i)$ is a random sample from $\mathcal{I}$.

The situation is simpler in the case of a jointly Gaussian distribution. In this case, all transformed variables follow univariate Gaussian distributions, even though one of the new variables is proportional to the centroid variable. 

A back-transformation from the variables $y_i$ to the original coordinates $x_i$ via the inverse  $M_i^{-1}$ then yields the desired correlated photon events. In all cases, we verified that the position distribution of the obtained random numbers agree to the original respective position space distribution functions.

\begin{figure}[t!]
\includegraphics[width=0.8\linewidth ]{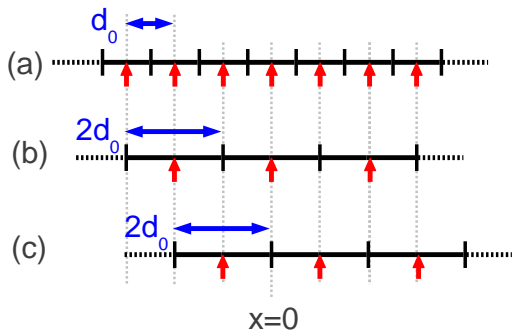}
\caption{\label{detection}(Color online) Model for the detection setup. (a) shows an array of detectors of size $d_0$. The red arrows indicate the central positions of the individual detectors. (b) shows the corresponding array with double detector size $d_0$. Compared to (a), only half of the possible detection positions occur. (c) shows the array of (b) suitably shifted that now the possible detection events cover those of (a) which are missing in (b).}
\end{figure}

\subsection{Modeling of the detection system}
\label{discretization}

The positions of the correlated $N$-photon states obtained in the previous section are continuously distributed. Any measurement, however, employs detectors of finite size, which yield a discretized position information.  To model this discretization, we assume an array of identical detectors, as shown in Fig.~\ref{detection}(a). Each detector has size $d_0$. We assume that the detectors do not overlap, and that there is no space between two adjacent detectors. Thus, each photon position of the continuous position distribution obtained from the random number generation can uniquely be assigned to a single detector. The numerical discretization procedure thus amounts to replacing the position of each individual photon with the central position of the detector which registers the respective photon. 

In a suitable coordinate system, the possible measurement outcomes are $i\,d_0$, with $i\in \mathbb{Z}$. The possible outcomes for the centroid coordinate in a $N$-photon detection with photons hitting the  $i_n$th detector ($1\leq n\leq N$) then follow from  Eq.~(\ref{centroid-def}) as
\begin{equation}
\label{discrete-centroid}
 X= \frac{d_0}{N}\, \sum_{n=1}^N i_n \,.
\end{equation}

If the detector size is changed, also the set of possible detection outcomes is changed. For example, in 
Fig.~\ref{detection}(b), detectors of double the size compared to those in (a) lead to a set of possible measurement outcomes which contain only every second possibility compared to that of (a). 
In addition, the detection array can be shifted, with an example for shift $d_0$ shown in Fig.~\ref{detection}(c). The measurements of (b) and (c) together allow for a set of potential measurement outcomes which coincides with that of (a). Note, however, that the detectors in (b) and (c) overlap spatially, such that the combined measurements of (b) and (c) are not equivalent to measurement (a). 

Throughout the later analysis, we will compare the performance with different detector sizes. In order to obtain comparable predictions for the wave function, we proceed as follows. For the smallest detector size $d_0$, we apply the centroid method for a single detection array position, which we denote as shift 0. Next, we double the detector size to $2d_0$, and obtain centroid data at the two shift positions 0 and $d_0$. Analogously, for detectors of size $m\,d_0$, we calculate centroid data for $m$ suitable shifts. This way, the wave function is estimated at the same set of positions for all detector sizes.

For this analysis, we employ two different methods for the calculations for different shifts. First, we generate $N_0$ realizations of the correlated $N$-photon state as described in Sec.~\ref{multivariates}. In method I, we then use the total number of $N_0$ events for each of the $m$ required shifts. In method II, we divide the total number of events $N_0$ by the number of shifts $m$, and evaluate each shift with $N_0/m$ events only. 
In terms of an experimental realization, in method I, the number of measurements required increases with detector size, whereas in method II, the number of measurements is independent of the detector size.

Due to the statistical nature of the measurement, in principle,  also a limited number of individual non-adjacent detectors  can be used, if their positions are shifted such that the combinations of all measurements cover the entire beam correlation area. This method has been adopted in~\cite{PhysRevLett.107.083603,Shin:10}.  While the measurement time increases, the complicated fabrication process for an array of detectors can be avoided in proof-of-principle measurements.


\subsection{Error estimation}
\label{error-rms}

To evaluate the performance of the centroid method for given detection parameters, we first fit the centroid data to the original distribution in order to obtain the optimum overall scaling factor for the centroid data, and multiply the centroid data with this factor. This is necessary, as we evaluate the centroid data only over a limited position range, such that the overall normalization of the measured centroid data a priori is unknown. We then employ the weighted root mean square deviation given by
\begin{equation}
\label{dev}
\frac{1}{\sqrt{b}}\sqrt{\sum_{i=1}^b||\psi(X_i)|^2-z_i|^2},
\end{equation}
where $b$ is the number of centroid data points, and  $|\psi(X_i)|^2$ and $z_i$ are the reference value from the original probability distribution and the estimate obtained from the centroid method, respectively, at positions $X_i$.

Note that the rms deviation in part depends on the spatial extent over which the centroid data is compared to the original wave function. If a larger range is considered, positions are included into the analysis for which both the original probability distribution and the centroid prediction are very low, such that they in essence do not increase the sum in the rms deviation, but only $b$. This is particularly important if rms values for different wave functions are compared.

\section{Results}
\label{results}

In the following, we present our results for three different trial states. We start with NOON states.
\subsection{NOON States}
\label{noon}
%
\begin{figure}[t!]
\includegraphics[width=0.8\linewidth ]{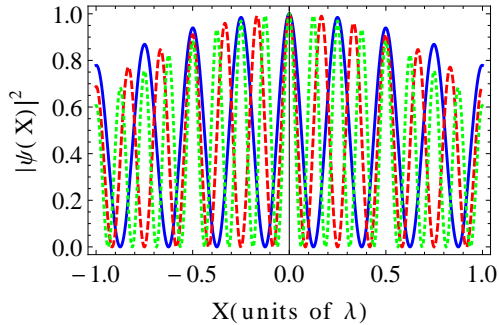}
\caption{\label{noon-a}(Color online) Result of the centroid method for NOON states with different $N$, shown against the centroid position in units of the single-photon transverse wavelength $\lambda$. Continuous blue, dashed red, and  dotted green curves correspond respectively to $N=2$, $3$ and $4$. The increase in the number of fringes clearly shows the resolution enhancement with $1/N$.}
\end{figure}
%

\subsubsection{Resolution enhancement}
Fig.~\ref{noon-a} depicts the the centroid probability distributions for two, three and four photon NOON states, obtained with a detector size small enough to resolve all features of the wave function. It can be seen that there are $2N$ fringes per wavelength for $N$ photons, confirming the expected scaling of the obtained resolution with $1/N$. We thus find that our simulation technique is able to recover the predicted resolution enhancement.

\subsubsection{\label{dep-size}Dependence on the detector size}
\begin{figure}[t]
\includegraphics[width=0.8\linewidth ]{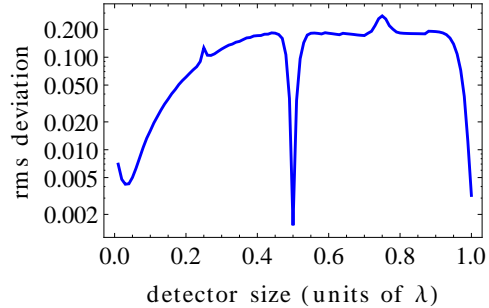}
\caption{\label{fig-rms-size}(Color online) Root-mean-square deviation of the recovered two-photon centroid distribution from the original wave function position space distribution as a function of the detector size. Note the logarithmic scale.}
\end{figure}

We thus turn to a detailed analysis of the centroid technique. As our first step, Fig.~\ref{fig-rms-size} shows the dependence of the rms deviation of the recovered two-photon centroid distribution from the original position space distribution as a function of the detector size. As expected, the deviation is low for small detectors, and initially increases with growing detector size. Starting from detector sizes of about $\lambda/2$, the rms deviation saturates. Next to this general structure, an increase of the deviation at detector size $\lambda/4$ and $3\lambda/4$, as well as sudden drops in the rms for $\lambda/2$ and $\lambda$ can be observed. Finally, towards very low detector sizes, the rms deviation starts to increase again. In the following, we will explain and interpret all of these features.

\begin{figure}[t]
\includegraphics[width=0.9\linewidth ]{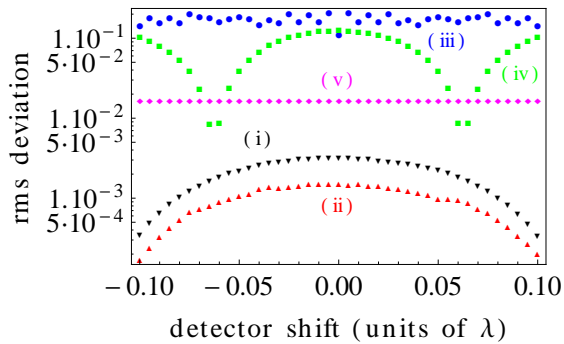}
\caption{\label{fig-rms-shift}(Color online) Root-mean-square deviation of the recovered two-photon centroid distribution from the original wave function position space distribution as a function of the detector shift. The different curves show results for the detector sizes (i) triangles faced right $\lambda$, {\color{red}(ii)} triangles faced left $\lambda/2$, {\color{blue}(iii)} blue blobs $0.3\lambda$, {\color{green}(iv)} green squares $\lambda/4$, and {\color{magenta}(v)} magenta rhombuses $\lambda/10$. Note the logarithmic scale.}
\end{figure}

First, we analyze the structures at $\lambda/4$, $\lambda/2$ and $\lambda$. For this, we choose a detector size, and then calculate the rms deviation as a function of a shift of the whole detector array along the centroid axis without modifying the detector size. The results are shown in Fig.~\ref{fig-rms-shift}. It can be seen that for detector sizes other than $\lambda/4$, $\lambda/2$ and $\lambda$, the rms deviation is approximately constant over the whole range of shifts. For these detector sizes, we conclude that the naive expectation that a smaller detector leads to a better recovery of the wave function is correct. 

In contrast, at detector size $\lambda/2$, the rms deviation strongly depends on the shift. Even more surprisingly, while there is a slight dependence on the detector shift for detector sizes $\lambda$ and $\lambda/2$, the rms deviation remains very low for all shifts despite the large detector size. We will show now that these result are artifacts of the calculation procedure, which arise due to the particular structure of the wave function to be recovered. 

\begin{figure}[t]
\centering
\includegraphics[width=0.7\linewidth ]{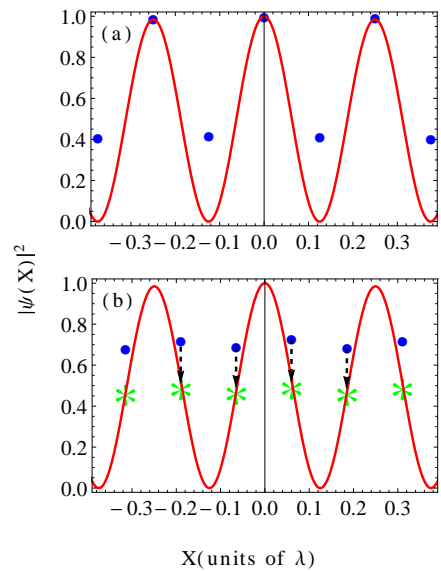}
\caption{\label{det-1/4}(Color online) Centroid analysis for detector size $\lambda/4$. The continuous red curve shows the probability distribution for a 2-photon NOON state. The blue dots indicate the results of a centroid measurement with particular detector settings. In (a), the detection array is not shifted with respect to the origin of the probability distribution, such that one of the centroid data points coincides with the center position $X=0$. (b) The detection array is shifted to $s=-13\lambda/200$, which corresponds to a minimum in the rms deviation in Fig.~\ref{fig-rms-shift}. The blue dots show the centroid data with the same scaling as in (a). The green asterisks show this data fitted to the original distribution. }
\end{figure}

\begin{figure}[t]
\includegraphics[width=0.7\linewidth ]{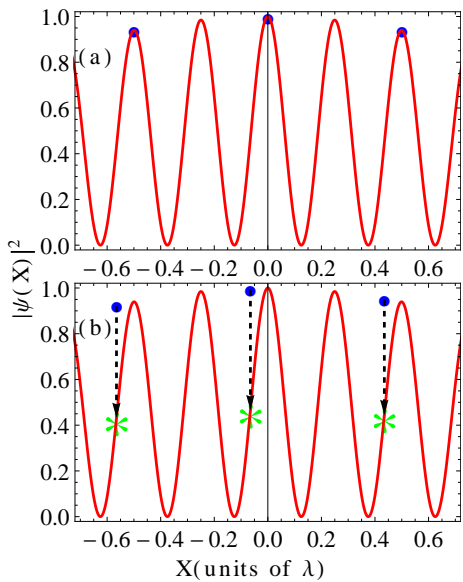}
\caption{\label{det-1}(Color online) Centroid analysis for detector size $\lambda$. The figure is analogous to Fig.~\ref{det-1/4}, except for the detector size.}
\end{figure}

Fig.~\ref{det-1/4} explains the situation for detector size $\lambda/4$. In (a), the original position distribution along the centroid direction for a two-photon NOON state is shown together with the centroid data obtained for detector size $\lambda/4$ and zero shift. For this shift value, one of the centroid data points coincides with the position distribution maximum at $X=0$. It can be seen that half of the centroid data points coincide perfectly with the original distribution - but the other half strongly deviates. The reason is that the position distribution is zero only at single points, whereas the large detectors cover a range around these zeros with non-zero photo detection probability. As a result, the centroid data is non-zero, in contrast to the probability distribution. This deviation for half of the points explains the large rms value at zero shift in Fig.~\ref{fig-rms-shift}. We next consider the same setting, but with shift $s=-13 \lambda/200$, corresponding to a minimum in the rms value for 
detector size $\lambda/4$ in Fig.~\ref{fig-rms-shift}. Fig.~\ref{det-1/4}(b) shows the corresponding centroid result obtained with the same scaling of the centroid data as in (a) as blue dots. It can be seen that the deviation to the probability distribution is strong. But if we fit the overall scaling factor of the centroid data to the original distribution, then good agreement is obtained. This is the reason for the minimum in Fig.~\ref{fig-rms-shift}, in which the centroid data is fitted to the original distribution for each shift position individually. 
The analogous analysis for detector size $\lambda$ is shown in Fig.~\ref{det-1}. In this case, due to the $\lambda$-periodicity of the wave function, a perfect fit of the centroid data is possible for any detector shift.

However, in an actual measurement, results from all shifts would have to be accounted for with the same overall prefactor, as an individual fitting to the unknown distribution to be determined is impossible. For this reason, detectors of size $\lambda$ or $\lambda/2$ cannot recover the wave function. To show this, we have taken a set of $N_0=10^6$ detection events, calculated the centroid distributions for a large number of shifts, and joined the corresponding distributions for the respective shifts without further processing them beforehand. Only the total set obtained from all shifts was then fitted in amplitude to the original position distribution. The result was a structure-less Gaussian which did not contain the sub-wavelength fringe features of the original distribution.  Therefore, in conclusion, it is not possible to recover the wave function with detectors of specific larger sizes, despite the small rms deviation for specific detector shifts, which has to be interpreted as an artifact of the 
analysis. Note, however, that such measurements with larger detectors could potentially be sufficient to analyze the periodicity or the symmetry of a given wave function.

\begin{figure}[t]
\includegraphics[width=0.9\linewidth ]{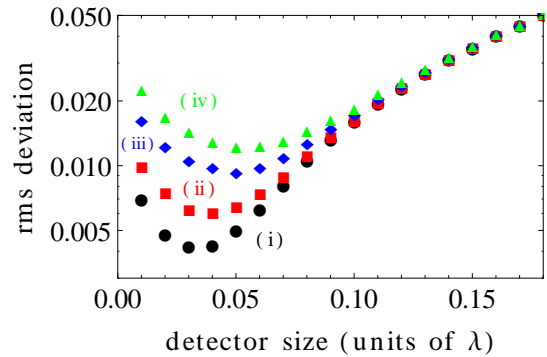}
\caption{\label{fig-small-det}(Color online) Effect of statistical fluctuations on the centroid methods for small detector sizes. The rms deviation was evaluated (i) black blobs, once for $10^6$ events, {\color{red}(ii)} red squares, twice for two subsets with $5\times10^5$ events, {\color{blue}(iii)} blue rhombuses, 5 times for subsets of $2\times 10^5$ events, and {\color{green}(iv)} green triangles, 10 times for subsets of $10^5$ events. In all cases, the same $10^6$ events were analyzed, and the respective subsets were chosen disjunct. The figure shows the results averaged over the respective subsets. Note the logarithmic scale.}
\end{figure}

Next, we analyze why the rms deviation increases again at very small detector sizes. We found that this is a statistical effect. If the detector size is decreased while the detection area is kept constant, the number of detectors increases. Then, the mean number of events per detector decreases with the detector size. At some point, the statistical fluctuations due to the decreasing number of events become large enough to dominate the rms deviation. In order to verify this interpretation, we evaluated the rms deviation (i) once for $10^6$ events, (ii) twice for two subsets with $5\times10^5$ events, (iii) 5 times for subsets of $2\times 10^5$ events, and (iv) 10 times for subsets of $10^5$ events. In all cases, the same $10^6$ events were analyzed, and the respective subsets were chosen disjunct. After this, we averaged the results over the respective subsets, such that in every case, the same events were analyzed. The result is shown in Fig.~\ref{fig-small-det}. It can be seen that the rms deviation of all 
cases approximately agrees towards larger detector sizes. But towards smaller detector sizes, the smaller the corresponding subsets are, the more the deviation increases. Note that also the difference in rms deviation between different subsets of equal size is much smaller than the difference in deviation between subsets of different size.

\subsubsection{\label{comparison}Comparison of the two methods}
\begin{figure}[t]
\includegraphics[width=0.9\linewidth ]{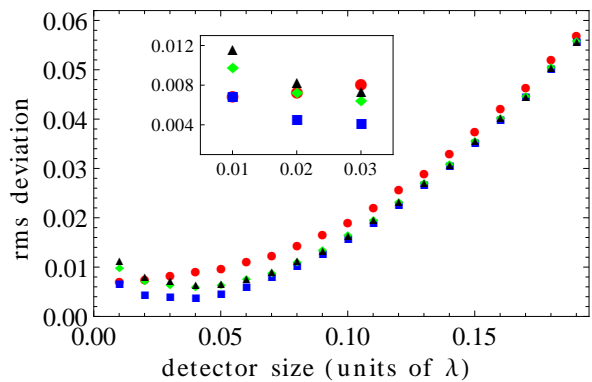}
\caption{\label{methods}(Color online) (a) Root-mean-square deviation as a function of detector size. Results are shown for a two-photon NOON state. The blue squares are obtained using method I discussed in the text, whereas the red dots are obtained using method II. The green diamonds and the black triangles are obtained using method II with one half and one third of the total number of events used in (a), respectively. Inset shows a magnification at small detector sizes.}
\end{figure}

In this section, we compare the two analysis methods I and II introduced in Sec.~\ref{discretization}. Fig.~\ref{methods} shows results for a 2-photon NOON state using the two methods. The two curves for method I (blue squares) and II (red dots) agree for larger detectors, but deviate for small detectors. In particular, the unexpected increase in the rms deviation towards low detector sizes observed in Fig.~\ref{fig-small-det} and interpreted as statistical fluctuations due to a low number of events per detection bin is absent for method II.

The reason for this qualitative difference is as follows. For the lowest detector size $d_0$, no shift is required, and the two methods are equivalent. Thus, the same rms deviation is obtained. 
As can be seen from the increase in the rms value for method I, the prediction at this detector size is already limited  by the low number of events per detection bin.
For the next higher detector size $2d_0$, in method I, the number of events per detection bin is increased, as the same number of $N_0$ events are distributed over a lower number of bins. In contrast, for method II, the number of events per bin is not increased, as with increasing detector size, the number of events considered for each detector shift is reduced. As a consequence, the rms deviation for method II monotonously increases with detector size, whereas that for method I can decrease with increasing detector size. In the latter case, the reduction of the rms due more events per bin on average outweighs the increase in rms due to the increase of the detector size. 

This interpretation is further supported  by the two other data sets in Fig.~\ref{methods}. These show method II, but with half the number of events  (green diamonds) or one third of the events (black triangles). It can be seen that the rms results for the smallest detector size $d_0$ roughly agree with those for method II with half the data points at detector size $2\,d_0$. Further, the rms for method II with one third of the data points agrees to this value at detector size $3\,d_0$. This shows that for such small detectors, the rms deviation is dominated by the low number of counts per bin.

We obtained qualitatively similar results also for $N\in\{3,4\}$.

\subsubsection{Multiphoton NOON states}
\label{multiphoton-noon-states}

\begin{figure}[t]
\includegraphics[width=0.95\linewidth ]{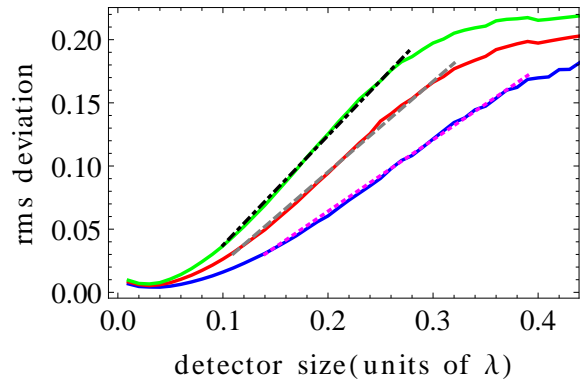}
\caption{\label{rmsallshifts-N}(Color online) Root-mean-square deviation versus detector size for NOON states using method I. Results are compared for number of photons $N=2$ (lower blue curve), $N=3$ (middle red curve), and $N=4$ (upper green curve). The lines indicate linear fits to the curves over the indicated detector size range.}
\end{figure}

%
We now turn to NOON states with higher photon number.  Fig.~\ref{rmsallshifts-N} shows the rms deviation as a function of the detector size for $N=2$, 3 and 4 photons.  Note that Fig.~\ref{rmsallshifts-N} has been generated including all detector shifts required for the respective detector sizes, such that the spurious maxima or minima in the rms deviation at specific detectors sizes  found in Fig.~\ref{fig-rms-size} due to the dependence on the shift position shown in Fig.~\ref{fig-rms-shift}  do not appear. All data points are generated using the same $N_0=10^6$ events with method I introduced in Sec.~\ref{comparison}.  
As the spatial extent over which the probability distribution is significantly larger than zero decreases with increasing photon number $N$, we adjust the position range over which the wave function is matched to the centroid measurement accordingly. Thus, the $3$- and $4$-photon cases are evaluated over $2/3$ and $2/4=1/2$ the range of that of the $N=2$ case, respectively. 

It can be seen that at small detector sizes, the rms deviation is low and approximately independent of $N$. In this limit, the detectors are chosen small enough to recover all features even of the $N=4$ wave function. With increasing detector size, the rms deviation starts differing from the small-detector limit value  first for the $N=4$, then for the $N=3$, and finally for the $N=2$ case. Increasing the detector size further, the rms deviation has an approximately linear dependence of the detector size, until it eventually saturates for the large-detector limit.

In the near-linear region, we estimate slopes $\alpha = \Delta$(rms deviation)/$\Delta$(detector size) for $N=2$, $3$ and $4$ as $0.57$, $0.72$ and $0.87$, as indicated in Fig.~\ref{rmsallshifts-N}, respectively. 
These values are consistent with a scaling with $\sqrt{N}$. One possible interpretation of this scaling could be as follows. If the detection positions of each of the $N$ photons acquires an uncertainty of order $d_0$ due to the finite detector size, then an estimate of the uncertainty of the centroid coordinate $X=\frac{1}{N}\sum_{i=1}^{N}x_i$ is given by $\delta X = d_0 / \sqrt{N}$. This estimate is motivated by the fact that the sum of $N$ normally distributed independent variables with widths $\delta x$ is again a normal distribution, with width $\sqrt{N}\delta x$. Approximately, it also holds for other distributions. Together with the prefactor $1/N$ in the centroid variable, a scaling of $1/\sqrt{N}$ in the centroid coordinate uncertainty is obtained. 
Next, we use that an increase of the detector size from the low-size limit leads to most significant contributions to the rms deviation in the regions of highest slope of the centroid probability distribution.  We denote such a centroid coordinate with highest slope as $x_{max}$, and note that the value of the $\cos^2()$ part in the probability distribution evaluates to $1/2$ at this point. We thus can estimate the scaling of the rms deviation via
\begin{align}
\frac{1}{\sqrt{\delta X}}
\sqrt{\int_{x_{max}-\delta X/2}^{x_{max}+\delta X/2} \left | \cos^2(2\pi N x) - \frac{1}{2}\right|^2 dx }\,.
\end{align}
To leading order in the detector size, this expression scales as $\sqrt{N}\,d_0$, i.e., linearly with the detector size, and with slope proportional to $\sqrt{N}$, as observed in the numerical data.
%
%
%

\subsubsection{Single- and multi-photon detection}

One motivation for the centroid measurement method is the technical difficulty to achieve  multi-photon or photon-number resolving detection, as required, e.g., for sub-wavelength correlated multiphoton measurement schemes. If the size of the individual detectors in the centroid method is small enough, the number of events in which two or more photons hit the same detector is negligibly small, such that no photon number resolution is required. 

Motivated by this, we have analyzed the percentage of two-photon events in our numerical calculations, as a function of detector size. We denote the position range over which the photon probability distribution is evaluated by $\rho$. In the numerical calculations, $\rho$ is chosen such that it encompasses all parts of the probability distribution which are significantly larger than zero. For detector size $d_0$, we then estimate the number of detection bins $p$ as integer closest to $\rho/d_0$.  In our numerical simulations, we have used $\rho=7\lambda$ for $N=2$. For $N>2$, we used accordingly $\rho=7\lambda \times 2/N$.

We then count those events in which all photons arrive in the same bin as a multi-photon detection event. There are $p$ different possibilities to realize a $N$-photon detection event. Overall, in our numerical calculation, there are $p^N$ different possible events, as we do not make use of the symmetrization of the photon states and therefore distinguish between events, e.g.,  in which two photons are detected at positions $x_1,x_2$ and $x_2, x_1$.
Thus, we expect a ratio of multi-photon events given by approximately $p/p^N$. In terms of the detector size, we find that  $p/p^N \sim d_0^{N-1}$.

We then numerically generate random detection events, discretize them to model the detection procedure, and count all events in which all photons arrive in the same detection bin as multi-photon detection events. Finally, we calculate the percentage of the such obtained multi-photon events out of all detection events. Our numerical analysis indeed confirms the analytical estimate. 

Note that for the NOON state, the ratio of two-photon events reaches about $14\%$ probability for detectors of size $\lambda$. 
This maximum value may appear high, but it arises from the fact that in our calculations the parameters are such that the non-negligible support of the probability distribution only covers few single-photon wavelengths. Then, a detector that is wavelength sized already covers a significant part of the total relevant centroid coordinate range. It is important to note that our calculations remain valid despite the two-photon events, since our numerical approach naturally is capable of photon-number resolving ``measurements'' such multiphoton events do not have to be discarded.

%
\begin{figure}[t!]
\includegraphics[width=0.8\linewidth ]{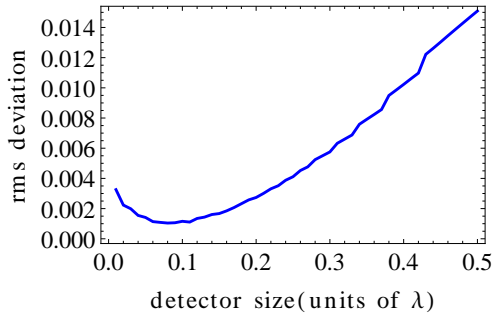}
\caption{\label{jnt-gaussian-rmsallshifts}(Color online) Root-mean-square deviation versus detector size for a jointly Gaussian state using method I. The parameters are same as Fig.~(\ref{fig-states}). }
\end{figure}

\subsection{Jointly Gaussian States}

We now turn to the analysis of jointly Gaussian states, as suggested in~\cite{centroid}. The analysis is performed in a similar way as for the NOON states. First results for the dependence of the rms deviation on the detector size incorporating all the shifts at larger detector sizes using method I are shown in Fig.~(\ref{jnt-gaussian-rmsallshifts}). As for the NOON state, the rms deviation increases with detector size, as naively expected. The increase of rms deviation towards the smallest considered detector sizes is again due to statistical effects, as explained in Sec.~\ref{dep-size}. Compared to the NOON states, the absolute values of the rms derivation are smaller, which is due to the much simpler structure of the Gaussian states. 
As for the NOON states, we also analyzed the rms deviation versus different detector sizes without taking into account shifts of the detector array. Contrary to the case of NOON states, there is no qualitative difference between the results with and those without shift. The reason is that for the Gaussian states, there is no fringe pattern in centroid direction which could match the periodicity of the detection array.

\subsubsection{Resolution Enhancement versus Multi photon Absorption}
\label{multiph-absorption}
%
%
\begin{figure}[t!]
\centering
\includegraphics[width=1\linewidth ]{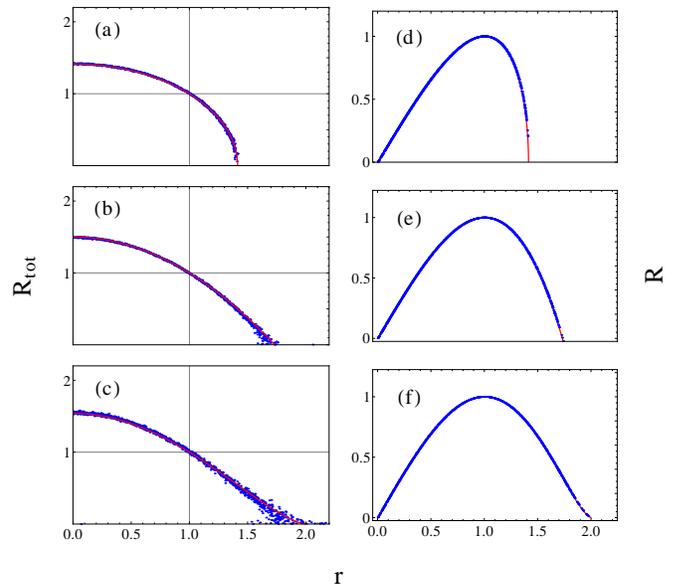}
\caption{\label{close-events-a}(Color online) $\langle k_n^2\rangle=\lambda^{-1}$. (a-c) Total multi photon absorption rate and (d-f) peak multi photon absorption rate versus the spot size reduction factor for jointly Gaussian state. (Top, middle and bottom row) correspond to $N\in\{2,3,4\}$, respectively. The data points obtained via numerical calculation are blue while red is the theoretical prediction. We multiplied the $R_{tot}$ values with $17.8,156,4100$ for $N\in\{2,3,4\}$, respectively to normalize the data such that at $r=1$ the corresponding value on the $y$-axis becomes unity too.}
\end{figure}

Next to the accuracy of the wave function recovery, also the efficiency is of relevance in any practical implementation. In essence, the measurement time is limited. This raises the question, how a limited number of measurements can be used in the most efficient way. In particular, in~\cite{multiphoton}, the efficiency of centroid detection was compared to that of a multiphoton detection scheme.
To analyze this in our numerical calculation, a practical way to distinguish single- from multi-photon detection events is required. We thus introduce a distance $d_{MP}=\lambda/400$, and denote $N$ photons which would hit a detector with unlimited position resolution at distance smaller than or equal to $d_{MP}$ a  $N$-photon event. Photons separated by more than $d_{MP}$ are registered as individual single-photon events. 

We start by analyzing the multi-photon absorption rate as a function of the spot size reduction factor $r$, which has been defined in~\cite{multiphoton} as the ratio of the classical width $W_C$ to the width $W$ of the probability distribution of the joint Gaussian defined by Eq.~(\ref{jtg}) for different choices of the parameters $B$ and $\beta$ such that $\langle k_n^2\rangle$ has a fixed value. Explicitly, $r=\sqrt{N}B/\sqrt{\langle k_n^2\rangle}$. $r=1$ defines the standard quantum limit, whereas the ultimate Heisenberg limit is given by $r=\sqrt{N}$. 

Our numerical scheme works as follows. We first fix $\langle k_n^2\rangle$ to a positive integer multiple of $\lambda^{-1}$. Then, we vary $B$ and calculate $\beta$ from Eq.~(\ref{knsq}). Note that if $\langle k_n^2\rangle$ is fixed to a large positive value in units of $\lambda^{-1}$, then many values of $r$ in the quantum regime given by $1<r\leq\sqrt{N}$ cannot be accessed because increasing $B$ results in a $\beta$ which makes $B$ less than its classical value $\beta/\sqrt{N}$ . Afterwards, we generate random detection events as explained in Sec.~\ref{multivariates}. Note that we do not apply discretization to mimic detection by finite size detectors. We then determine those events which contain photons with distance smaller than or equal to $d_{MP}$ and call these \emph{close events}. The width of the histogram of these close events provides us with the width $W$, which is calculated by fitting the histogram data to a Gaussian distribution of the form $c e ^{-d x^2}$. The width $W$ is then given by $2/(\
sqrt{2d})$.
Afterwards, we divide the number of close events by the total number of $N$-photon events to calculate the normalized total multiphoton absorption rate $R_{tot}$.

Fig.~(\ref{close-events-a}a) shows our results for the normalized total multiphoton absorption rate as a function of the spot size reduction factor  for $N\in\{2,3,4\}$. It can be seen that the results quantitatively agree with the theoretical prediction $R_{tot}=((N-r^2)/(N-1))^{(N-1)/2}$ found in~\cite{multiphoton}. Note that the statistical fluctuation of our numerical data increases with $r$ towards the limiting value $\sqrt{N}$. The reason for this is that once the variance of $k_n$ is fixed, an increase in $B$ is balanced by a decrease in $\beta$. At $r=R_{tot}=1$, the standard quantum limit is obtained which corresponds to a unit multi-photon absorption and a classical distribution. As $B$ is increased, the probability distribution becomes more narrow along the centroid direction, but expands along the orthogonal directions of the relative coordinates. This implies that the distance between any two random coordinates can be larger now. This is why with an increase in $B$, the number of \emph{close 
events} decreases. For this calculation, we extracted $1$ million pairs of random events. At $B=0.01\lambda^{-1}$, out of the 1 million events, some $79,201$ are \emph{close events} while at $B=0.999\lambda^{-1}$, this number reduces to $3316$. 

Using our $r$, we can calculate normalized peak multi photon absorption rate defined in~\cite{multiphoton} as $r((N-r^2)/(N-1))^{(N-1)/2}$. Fig.~(\ref{close-events-a}d) shows the corresponding result. The blue data points show the numerical results while red curve plots $r\sqrt{2-r^2}$. Again, the two results agree at most of the range of spot size reduction factor $r$ but $r=\sqrt{N}$ is not approached by the numerical data.

Note that we obtained qualitatively similar results also for $N=3$ and $N=4$ photon states, shown by Fig.~(\ref{close-events-a}b), (e) and (c), (f), respectively.

\subsubsection{Resolution Enhancement versus Multi photon Absorption for a Fixed Feature Size}
\begin{figure}[t!]
\includegraphics[width=0.8\linewidth ]{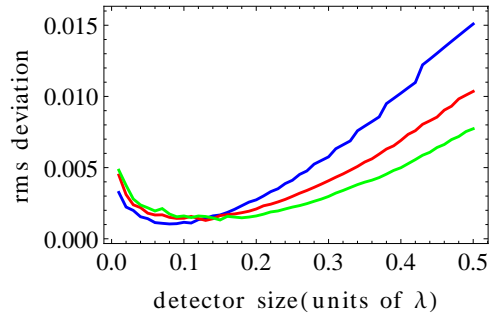}
\caption{\label{fixedsize}(Color online) Root-mean-square deviation plotted versus the detector size for a fixed physical size of the quantum mechanical jointly Gaussian centroid probability distribution for various $N$ using method I. The blue, red and green curves correspond respectively to $N=2$, $3$ and $4$. The parameters are explained in the text.}
\end{figure}
So far, the analysis has been limited to cases in which the feature size decreases with increasing number of detected photons $N$. Now, we compare the rms deviation as a function of detector size for different $N$, but with physical feature size kept constant. 
For this, we notice that the probability distribution in centroid direction scales as $\exp(-2 N^2 B^2 X^2)$, with $X$ the centroid coordinate. For each $N$, we choose $B$ in such a way that the resulting spatial extent of the distribution becomes $\exp(- 8  (X/\lambda)^2)$, which leads to $B = 2/(N\lambda)$.
We further choose $\beta = 1/\lambda$ for $N=2,3$. For $N=4$, we instead choose $\beta = 4/5\,/\lambda$, such that in all cases the parameters fulfill $B>\beta /\sqrt{N}$ and thus the probability distribution is non-classical.


The results presented in Fig.~(\ref{fixedsize}) show that as $N$ is increased, the value of rms deviation for a fixed detector size is decreased for most of the range of detector sizes. This is to be expected because if the physical size of the distribution is fixed, absorbing more photons should improve the result of the measurement. As explained before, the results at very small detector sizes are prone to numerical fluctuations if a fixed number of events is distributed over more and more detector bins. 

Qualitatively similar results have also been obtained using method II. Again we found that method II leads to slightly higher rms deviations in particular at lower detector sizes, because of the lower number of events included in the calculation.

\subsection{Correlated Coherent Cat States}
\label{corr-coh-cat}

\begin{figure}[t!]
\includegraphics[width=0.8\linewidth ]{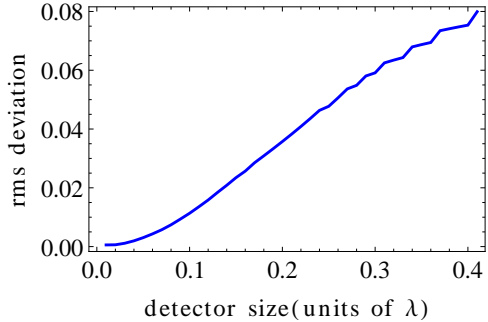}
\caption{\label{cat1-rmsallshifts}(Color online) Root-mean-square deviation against detector size for a correlated coherent cat state with $\alpha=\imath$. Results have been obtained using method I. }
\end{figure}

Finally, we turn to coherent cat states. As for the other states, we start by discussing the rms deviation against the detector size. Results are shown in Fig.~\ref{cat1-rmsallshifts}. Note that this figure was obtained including all different shift positions for the larger detectors, using method I such that the positions at which centroid data is obtained are the same for all detector sizes. The results is as expected, with a certain range of small detector sizes over which the recovery of the wave function is good, followed by a continuous increase of the rms with the detector size, until the rms saturates towards large detector sizes.

\begin{figure}[t!]
\includegraphics[width=0.8\linewidth ]{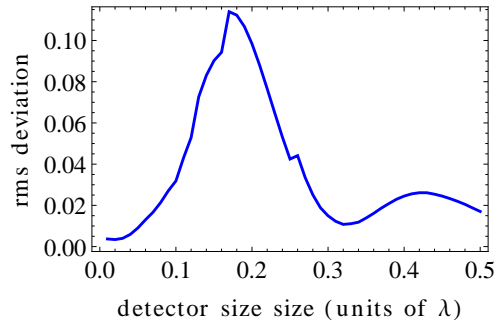}
\caption{\label{cat1-rms-shift0}(Color online) Root-mean-square deviation versus detector size for a correlated coherent cat state with $\alpha=\imath$. In contrast to Fig.~\ref{cat1-rmsallshifts}, results for a single fixed detector position are shown.
}
\end{figure}

In contrast, the rms deviation for single fixed detection arrays of different sizes show a rather different behavior compared to the other considered states, as shown in Fig.~\ref{cat1-rms-shift0}. Here, an initial increase in rms deviation with the detector area is seen, as expected, but for detector sizes of about $0.17\lambda$, the rms deviation acquires a maximum and afterwards oscillates with the detector size. These features again can be explained via the shape of the probability distribution of the original wave function. Results for other shifts of the detection array give similar results for detector sizes up to about $\lambda/10$, but afterwards deviate significantly.

\begin{figure}[t!]
\includegraphics[width=0.45\linewidth ]{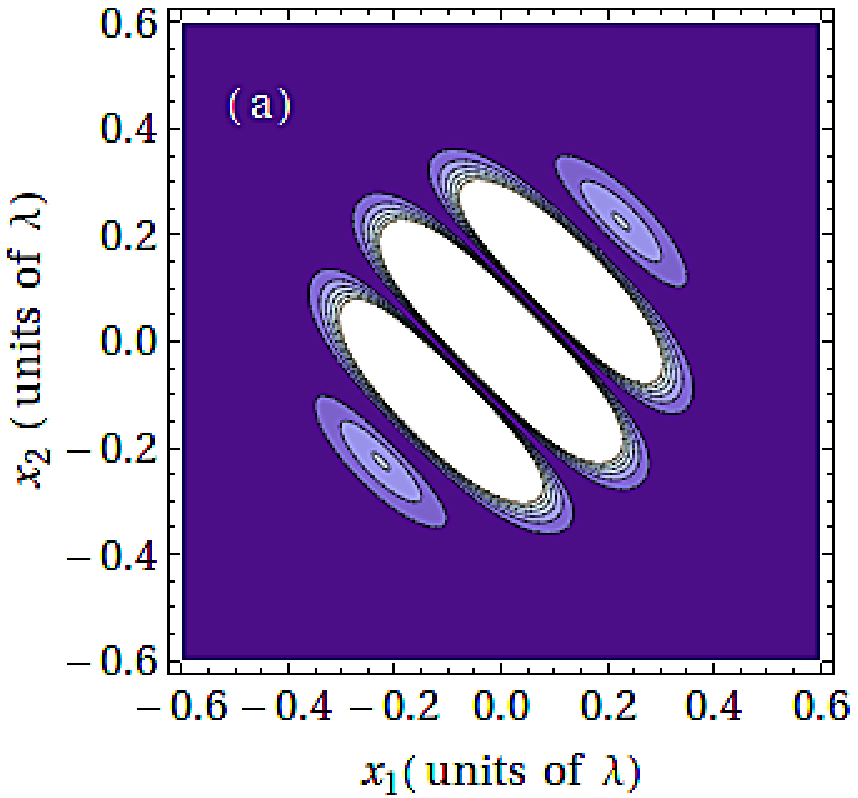}
\includegraphics[width=0.45\linewidth ]{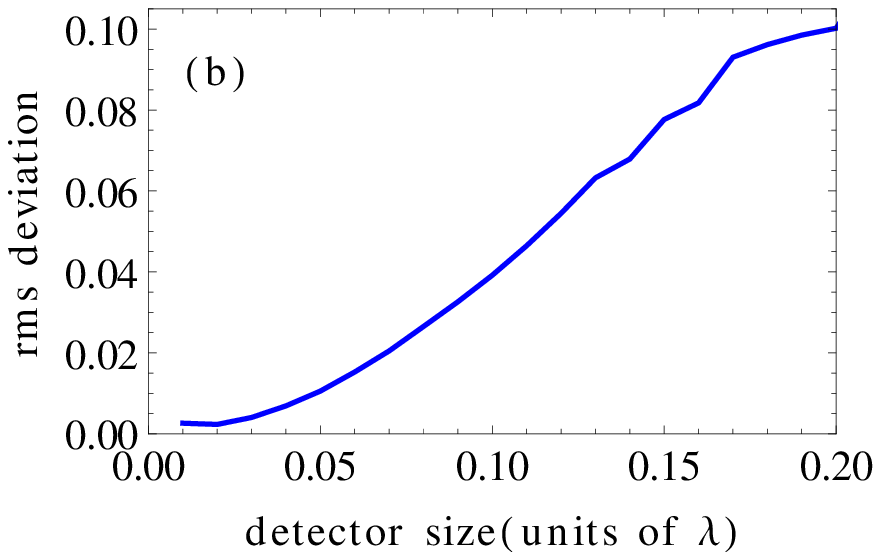}
\caption{\label{cat-2}(Color online) (a) Probability distribution  for a correlated coherent cat state with $\alpha=\imath\sqrt{2}$ in position space. (b) Root-mean-square deviation against detector size using method I.}
\end{figure}

As can be seen from Eq.~(\ref{psisq-cat-scale}), cat states have the advantage that the fringe pattern in centroid direction superimposed onto the overall Gaussian wave function envelope has a periodicity which can be controlled by the magnitude of $\alpha$. Since $|\alpha|^2$ is the mean number of photons in the corresponding cat state, this change in the fringe pattern goes along with a change in the number of photons in the light field. But nevertheless, some of the realizations of the cat state will consist of two photons independent of $|\alpha|$, such that we can continue to evaluate only those detection events. In this way, the feature size can be controlled without changing the number of detected photons.  An example for a correlated coherent cat state with higher mean photon number ($\alpha=\imath\sqrt{2}$) is shown in Fig.~(\ref{cat-2}). As compared to Fig.~\ref{fig-states}(f) with lower mean photon number, the number of fringes in the position space probability distribution increased as expected. 
Note that 
a 
similar shrinking was observed in Fig.~\ref{noon-a} for NOON states with higher number of detected photons. Here, in contrast, the number of detected photons is kept constant at 2. 
%
%
%
%
%

\begin{figure}[t!]
\includegraphics[width=0.8\linewidth ]{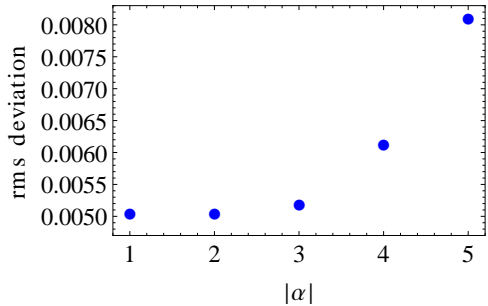}
\caption{\label{absorption}(Color online) Root-mean-square deviation for correlated cat states as a function of $|\alpha|$. In all cases, two-photon detection events are considered.  $N_0=10^5$ events were considered with detector size $\lambda/100$. }
\end{figure}
%
Fig.~\ref{absorption} shows the rms deviation as a function of $|\alpha|$. With increasing $|\alpha|$, the feature size of the wave function to be recovered becomes smaller, such that for a fixed detector size, increasing rms deviation with increasing $|\alpha|$ is expected. 
One can notice that the rms deviation initially remains approximately independent of the magnitude of $|\alpha|$, and then increases with growing $|\alpha|$. We can thus conclude that the parameters chosen such that for $|\alpha|$ below approximately $2.5$ the detector size is small enough to recover the wave function as good as possible, limited by the statistical uncertainty due to the finite number of detection events. Higher $|\alpha|$ lead to feature sizes which cannot be recovered fully.

\begin{figure}[t!]
\includegraphics[width=0.8\linewidth ]{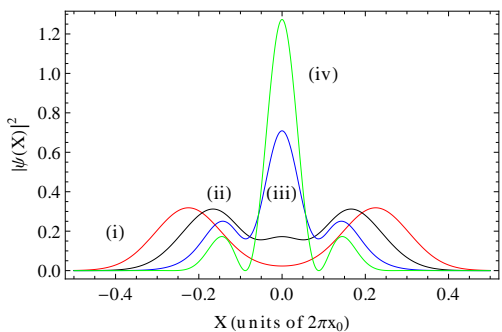}
\caption{\label{fig-phi}(Color online) Position space probability density along the centroid coordinate for $|\alpha|=1$ and (i) $\phi = 0$, (ii) $\phi = \pi/8$, (iii) $\phi = 3\pi/8$, (iv) $\phi = \pi/2$. }
\end{figure}

Finally, we exploit that the coherent cat states allow one to continuously tune the shape of the centroid position space probability density from a periodic fringe pattern to less periodic structures. This can be achieved, e.g., by choosing different phases $\phi$, as shown in Fig.~\ref{fig-phi}. For $\phi = 0$, an approximate double-peak structure is achieved. $\phi = \pi/2$ leads to the multi-fringe pattern studied so far, with relative heights of the peaks given by the overall envelope of the wave function. But for intermediate values, the shape of the wave function can be changed, such that, for example, the central peak is suppressed ($\phi = \pi/8$. This way, the shape of the probability density can be modified from the periodic fringe pattern observed, e.g., for NOON states. This is significant not least due to the dependence of the rms deviation on the structure of the wave function we observed in Sec.~\ref{dep-size}. We performed the centroid analysis for a number of different values for $\phi$ 
ranging from $0$ to $\pi/2$ and found that the rms deviation as a function of the phase $\phi$ has some residual fluctuations which are compatible with a constant value independent of the choice of $\phi$.  It is interesting to note that the position space probability density as a function of the two position variables $x_1$ and $x_2$ is rather different for $\phi=\pi/2$ than for $\phi=0$. For $\phi = 0$, it consists of two near-circular structures. But for $\phi=\pi/2$, the probability density is strongly elongated along the direction perpendicular to the centroid axis, and narrowly-spaced fringes along the centroid axis. Thus we conclude that the centroid method is suitable to recover non-periodic patterns along the centroid axis, independent of the precise structure of the position space probability density along the directions perpendicular to the centroid axis.

\section{Summary }
\label{summary}

In summary, we have analyzed the optical centroid method for superresolution lithography using numerical experiments. In the respective limits, our results fully agree with the  predictions from the previous approximate analytical analysis. But our numerical treatment allows us to gain additional insight, also beyond the validity range of the analytical predictions. We exploited this mainly by studying the different observables in detail as a function of the detector size. A first analysis revealed unexpected features in the root-mean-square deviation of the recovered wave function compared to the true wave function. In the limit of low detector size, the rms does not continuously decrease as expected, but eventually increases again. This could be traced back to fluctuations in the number of events per detector bin which grows with decreasing detector size if the number of events is kept constant. In addition, a fixed detection array can lead to spurious maxima or minima in the rms deviation at specific 
detector sizes. In particular, we found very low value of rms deviation for large detectors of certain sizes. These features are found to originate from the structure of the wave function, but in general do not allow to recover the wave function with high resolution using large detectors. Such effects, however, could be used to detect certain properties of wave functions such as symmetries, or to efficiently distinguish between different wave functions using an array of large detectors.
We then augmented this analysis with results for NOON states with higher photon number, and with results for jointly Gaussian states with fixed feature size, but different photon number. Interestingly, for the NOON states, we found a regime of intermediate detector sizes in which the rms deviation depends approximately linearly on the detector size. One plausible interpretation of this effect is in terms of the error in estimating the centroid coordinate arising from the non-zero detector size, evaluted at the part of the wave function with highest slope which contributes most to the rms deviation. In the final part, we showed that the OCM can also be applied to cat states made out of superpositions of coherent states. The coherent cat states have the interesting property that their spatial feature size can be continuously tuned via the magnitude of the coherent state parameter $\alpha$.  Also the structure of the cat state position space probability density along the centroid axis can be modified via the 
phase of $\alpha$. This way, for example, shapes other than the periodic fringe pattern observed, e.g., for NOON states, can be probed. We found that the centroid method works equally well for different structures of the cat state. Since the number of photons in the cat state input pulse is not fixed, we focused on detection events with a certain number of photons. In this sense, our results in this section also relate to post-selection methods.

\acknowledgments

QG acknowledges support from the Higher Education Commission (HEC) of Pakistan administered by Deutscher Akademischer Austauschdienst (DAAD), from the International Max Planck Research School (IMPRS) for Quantum Dynamics in Physics, Chemistry and Biology, Heidelberg, Germany and from the Heidelberg Graduate School for Fundamental Physics (HGSFP).

%

\end{document}